\documentclass[onecolumn,showpacs,preprintnumbers,amsmath,amssymb]{revtex4}
                                                                                
\usepackage{graphicx}
\usepackage{dcolumn}
\usepackage{subfigure}
\usepackage{bm}
\newcommand{\chii}{\tilde{q}}
\def\be{\begin{equation}}
\def\ee{\end{equation}}
\def\bc{\begin{center}}
\def\ec{\end{center}}
                                                                                
\begin{document}
                                                                                
\preprint{SINP/06/2009}
                                                                                
\title{Quantum phase transition in a disordered
long-range transverse Ising antiferromagnet}
                                                                                
\author{Anjan Kumar Chandra$^{1}$}
\email{anjan.chandra@saha.ac.in}
\author{Jun-ichi Inoue$^{2}$}
\email{j_inoue@complex.eng.hokudai.ac.jp}
\author{Bikas K. Chakrabarti$^{1}$}
\vskip0.1cm
\email{bikask.chakrabarti@saha.ac.in}
\affiliation{%
\vskip0.2cm
$^{1}$Centre for Applied Mathematics and Computational Science and
Theoretical Condensed Matter Physics Division, Saha Institute of Nuclear 
Physics, 1/AF Bidhannagar, Kolkata-700064, India.\\
\vskip0.1cm
$^{2}$Complex Systems Engineering, Graduate School of Information
Science and Technology, Hokkaido University,
N14-W9, Kita-ku, Sapporo 060-0814, Japan\\
}%

\date{\today}
                                                                                
\begin{abstract}
We consider a long-range Ising antiferromagnet put in a transverse 
field (LRTIAF) with disorder. We have obtained the phase diagrams for both
the classical and quantum case. For the pure case applying quantum Monte Carlo 
method, we study the variation of
order parameter (spin correlation in the Trotter direction), susceptibility
and average energy of the system for various values of
the transverse field at different temperatures. The antiferromagnetic
order is seen to get immediately broken as soon as the thermal or
quantum fluctuations are added. We discuss generally the phase diagram for
the same LRTIAF model with perturbative Sherrington-Kirkpatrick (SK) type 
disorder. We find that while the antiferromagnetic order
is immediately broken as one adds an infinitesimal transverse field or thermal 
fluctuation to the pure LRTIAF system, an infinitesimal SK spin glass disorder 
is enough to
induce a stable glass order in the LRTIAF. This glass order
eventually gets destroyed as the thermal or quantum fluctuations are increased 
beyond their threshold values and the transition to para phase occurs. 
Analytical studies for the phase transitions are discussed in detail in each 
case. These transitions
have been confirmed by applying classical and quantum Monte 
Carlo methods. We show here that the disordered LRTIAF has a surrogate 
incubation property of the SK spin glass phase.
\end{abstract}

                                                                                
\def\be{\begin{equation}}
\def\ee{\end{equation}}
\maketitle

\section{Introduction} 

\noindent Quantum phases in frustrated systems are being intensively 
investigated these
days; in particular in the context of quantum spin glass and quantum 
axial next-nearest-neighbor Ising (ANNNI)
models \cite{Bhatt,CDS,Young,Ma,Canning,Kim}. Here we study in general the 
long-range Ising antiferromagnet put under transverse field (LRTIAF) with 
disorder in 
co-operative interactions superposed 
on it. We study here extensively, both analytically as well as numerically, the 
phase diagram for this model. As a special case, we also studied the pure 
long-range transverse Ising antiferromagnet model (i.e. no disorder). 

The finite temperature properties of 
sub-lattice decomposed version of the pure limit of this model was already
considered earlier \cite{Chakrabarti,Das}. The quantum phase transition and
entanglement properties of the full long-range model at zero temperature has 
also been studied \cite{Vidal}. 

Here we
present some results obtained by applying analytical as well as Monte Carlo
techniques \cite{Suzuki} to the general full long-range model at finite
temperatures and transverse fields. We observe indications of a very unstable 
quantum 
antiferromagnetic (AF) phase ($50\%$ spin up, $50\%$ spin down, without any
sub-lattice structure) in the pure LRTIAF
model, where the antiferromagnetically ordered phase gets destabilized by both
infinitesimal thermal (classical) as well as quantum
fluctuations (due to tunneling or transverse field) and the system becomes
disordered or goes over to the para phase \cite{Chandra}. This kind of
phase transition has also been studied by perturbative treatment 
\cite{Ganguli}.

When a little spin glass-like
disorder is incorporated with this pure LRTIAF model, the frustration is seen 
to destabilise the AF phase and stabilise a spin glass order.
To check how this `liquid'-like antiferromagnetic phase
of the pure LRTIAF gets `frozen' into spin-glass phase when a little disorder 
is
added, we study in general the LRTIAF Hamiltonian with a coupling with the 
SK spin glass Hamiltonian and study this entire system's phase transition 
behaviour
induced by both thermal and tunneling field. Indeed, stable SK like spin glass 
phase is observed for both thermal or quantum fluctuations below finite 
threshold values \cite{Chandra}.

This paper is organized in the following manner.
In Section II, we introduce the quantum LRTIAF model with SK disorder.
Then in the four subsections we discuss the analytical studies and simulations
for some special cases of this general model.
In Section II.A., we consider the quantum LRTIAF model without disorder, in
Section II.B. the classical model (i.e., without the transverse field), in
II.C. the quantum model at finite temperature and in II.D. the quantum model
at zero temperature. The detail calculation of free-energy is given in 
Appendix A and an exact analysis at T=0 is given in Appendix B.
In Section IV, we present some discussions on our results.

\section{LRTIAF with SK disorder}

\noindent The general model we study here is given by the following Hamiltonian
\begin{eqnarray}
H & = & - \frac{1}{N}
\sum_{ij (j > i) }
(J_{0}+\tilde{J}\tau_{ij})
\sigma_{i}^{z}
\sigma_{j}^{z} - h\sum_{i=1}^N \sigma^z_i -
\Gamma \sum_{i}\sigma_{i}^{x},
\label{eq:Hamiltonian}
\end{eqnarray}
where $J_{0}$ is the parameter controlling the strength of the 
antiferromagnetic bias and $\tilde{J}$ is an amplitude of the disorder
$\tau_{ij}$ in each pair interaction. $h$ and $\Gamma$ denote respectively the longitudinal and transverse
fields. The $\Gamma$ controls the 
quantum-mechanical fluctuation. 
Here $\sigma^x$ and $\sigma^z$ denote the $x$ and $z$ component of the $N$
Pauli spins
\[ \sigma^{z}_i = \left( \begin{array}{cc}
1 & 0 \\
0 & -1 \\ \end{array} \right); \hspace*{0.5cm}
\sigma^{x}_i = \left( \begin{array}{cc}
0 & 1 \\
1 & 0 \\ \end{array} \right); \hspace*{0.5cm} i = 1,2,....,N.
\]
As such the model has a fully frustrated (infinite-range or infinite dimensional) co-operative term.
When we assume that the disorder $\tau_{ij}$ 
obeys a Gaussian with mean zero and variance unity, the new variable 
$J_{ij} \equiv J_{0}+ \tilde{J}\tau_{ij}$ follows the Gaussian distribution,
$P(J_{ij}) =  {\exp}[-{(J_{ij}-J_{0})^{2}}
/{2\tilde{J}^{2}}]/{\sqrt{2\pi}\tilde{J}}$.
Therefore, we obtain the `pure' antiferromagnetic Ising model with infinite 
range interactions when we consider the limit $\tilde{J} \to 0$ keeping
$J_{0} < 0$. Of course the model with $J_{0} > 0$ and $\Gamma =0$, is identical 
to the classical SK model and with $J_{0} < 0$ and $\Gamma =0$ it is the 
LRIAF model.

For an analytic (mean field) study of the model we define an effective 
magnetic field $\vec{h}_{eff}$ at each site, which is 
a resultant of the average cooperation enforcement in the $z$-direction and the applied transverse field in the $x$-direction, so that the above Hamiltonian 
can be written as,\\
\be
H = \vec{h}_{eff}.\sum_{i=1}^N \vec{\sigma}_i ,
\ee
where
\be
\vec{\sigma}_i = \sigma^z_i \hat{z} + \sigma^x_i \hat{x}, \notag
\ee
and
\begin{subequations}
\begin{eqnarray}
\vec{h}_{eff} & = & (\vec{h}_{eff})^z \hat{z} + (\vec{h}_{eff})^x \hat{x} \notag \\
\mbox{} & = & \left( h + J_0 m^z + \tilde{J}\sqrt{q}y \right) \hat{z} + \Gamma \hat{x} , \\
|\vec{h}_{eff}| & = & \sqrt{(h + J_0 m^z + \tilde{J}\sqrt{q}y )^2 + \Gamma^2} ~~~ .
\label{eq:Hamiltonian3}
\end{eqnarray}
\end{subequations}
This replacement of $\sigma^z_j$ by its average value $\langle \sigma_{j}^{z} \rangle \equiv m^z$ in 
$(\vec{h}_{eff})^z$ should be valid for this infinite range model 
(see Appendix A for its much more precise description 
under replica symmetric theory).
The Gaussian distributed random field $\sqrt{q}y$ comes from the
local field fluctuation (see e.g., \cite{Binder}) given by the spin glass order parameter
(see Appendix A for details).
The average magnetisation is then given by
\begin{eqnarray}
\vec{m} & = & \frac {Tr \vec{\sigma} e^{-\beta H}}{Tr e^{-\beta H}} \notag \\ & = & (\tanh \beta |\vec{h}_{eff}|).\frac{\vec{h}_{eff}}{|\vec{h}_{eff}|}
\end{eqnarray}
and hence we have \\
{\small
\begin{eqnarray}
m^z & = & 
\int_{-\infty}^{\infty}
Dy 
\frac{J_{eff}}
{\sqrt{J_{eff}^{2}+
\Gamma^{2}}}
\tanh \beta 
\sqrt{J_{eff}^{2}+
\Gamma^{2}} 
\label{eq:quantum_mz} \\
m^x & = & 
\int_{-\infty}^{\infty}
Dy 
\frac{\Gamma}
{\sqrt{J_{eff}^{2}+
\Gamma^{2}}}
\tanh \beta 
\sqrt{J_{eff}^{2}+
\Gamma^{2}} 
\label{eq:quantum_mx} 
\end{eqnarray}}
{\small
\be
q =   
\int_{-\infty}^{\infty}
Dy 
\left\{
\frac{J_{eff}}
{\sqrt{J_{eff}^{2}+
\Gamma^{2}}}
\right\}^{2}
\tanh^{2} \beta
\sqrt{J_{eff}^{2}+
\Gamma^{2}}~~,
\label{eq:quantum_q} 
\ee}
\noindent where $J_{eff} = h + J_{0}m^z + \tilde{J}\sqrt{q}y$ and
$m \equiv N^{-1}\sum_{i}\langle \sigma_{i}^{z} \rangle$ is the magnetization 
and $q \equiv N^{-1}\sum_{i}\langle \sigma_{i}^{z} \rangle^{2}$
is the spin glass order parameter. We defined
$Dy \equiv dy\,
{\rm e}^{-y^{2}/2}/
\sqrt{2\pi}$. 
In Appendix A, we confirm that the above mean-field equations  are 
identical to the results obtained by the replica symmetric theory at the ground state
($\beta = \infty$). 

For the antiferromagnetic ($J_0 < 0$) and/or the spin glass phase (with 
$h = 0$), $m^z = 0$ is the only solution. We then have
{\small
\be
m^x =  
\int_{-\infty}^{\infty}
Dy 
\frac{\Gamma}
{\sqrt{(\tilde{J}\sqrt{q}y)^{2}+
\Gamma^{2}}}
\tanh \beta 
\sqrt{(\tilde{J}\sqrt{q}y)^{2}+
\Gamma^{2}}~~, \ee}
{\small \be
q =   
\int_{-\infty}^{\infty}
Dy 
\left\{
\frac{\tilde{J}\sqrt{q}y}
{\sqrt{(\tilde{J}\sqrt{q}y)^{2}+
\Gamma^{2}}}
\right\}^{2}
\tanh^{2} \beta 
\sqrt{(\tilde{J}\sqrt{q}y)^{2}+
\Gamma^{2}}~~.
\ee}

\subsection*{\bf II.A.  Pure LRTIAF  model}
\noindent The Hamiltonian of the infinite-range quantum Ising
antiferromagnet (without any spin glass disorder) is
\begin{eqnarray}
H & \equiv & H^{(C)} + H^{(T)} \notag \\ & = & - \frac{J_0}{N}\sum_{i,j(>i)=1}^N\sigma^z_i \sigma^z_j  
-  h\sum_{i=1}^N \sigma^z_i - \Gamma \sum_{i=1}^N \sigma^x_i ,
\label{eq:Hamiltonian1}
\end{eqnarray}
where $J_0$ denotes the long-range antiferromagnetic $(J_0<0)$ exchange 
constant.
We have denoted the co-operative term of
{$H$} (including the external longitudinal field term) by $H^{(C)}$ and the
transverse field part as $H^{(T)}$.
As such the model has a fully frustrated (infinite-range or infinite dimensional) co-operative term. 

\subsubsection*{\bf II.A.1. Analytical studies}
\noindent For $h = 0$, the Hamiltonian can be written as
\begin{eqnarray}
H & = & - \frac{J_0}{N}\sum_{i,j(>i)=1}^N\sigma^z_i \sigma^z_j - \Gamma \sum_{i=1}^N \sigma^x_i \notag \\ & = & - \frac{J_0}{N}\left( \sum_{i=1}^N \sigma^z_i\right)^2 - \Gamma \sum_{i=1}^N \sigma^x_i ~~.
\label{eq:Hamiltonian2}
\end{eqnarray}
Without the disorder term, the $\vec{h}_{eff}$ in Eq. 3(a) gets
modified to, \\
\be
\vec{h}_{eff} = J_{0}m^z \hat{z} + \Gamma \hat{x} ~~.
\ee
With this modified field, the expressions of
$m^z$ and $m^x$ become\\
\begin{subequations}
\begin{eqnarray}
m^z & = & \left (\tanh \beta \sqrt{(J_0 m^z)^2 + \Gamma^2}\right)\frac{J_0 m^z}{\sqrt{(J_0 m^z)^2 + \Gamma^2}}\\
m^x & = & \left (\tanh \beta \sqrt{(J_0 m^z)^2 + \Gamma^2}\right)\frac{\Gamma}{\sqrt{(J_0 m^z)^2 + \Gamma^2}}~~.
\end{eqnarray}
\end{subequations}
When $J_0 < 0$, then $m^z = 0$ is the only solution of Eq. (13a).
At zero temperature and at zero longitudinal and
transverse fields, the
$H^{(C)}$ would prefer the spins to orient in $\pm z$ directions only with
zero net magnetization in the $z$-direction. This antiferromagnetically
ordered state is
completely frustrated and highly degenerate. Switching on the transverse field
$\Gamma$ would immediately induce all the spins to orient in the
$x$-direction (losing the degeneracy), corresponding to a maximum of the
kinetic energy term and this discontinuous
transition to the para phase occurs at $\Gamma = 0$ (see Appendix B for an 
exact result at T = 0). However, at any
finite temperature the entropy term coming from the extreme degeneracy of the
antiferromagnetically ordered state and the close-by excited states does not
seem to
induce a stability of this phase.
\begin{figure}
\bc
\noindent \includegraphics[clip,width= 9cm, angle = 0]
{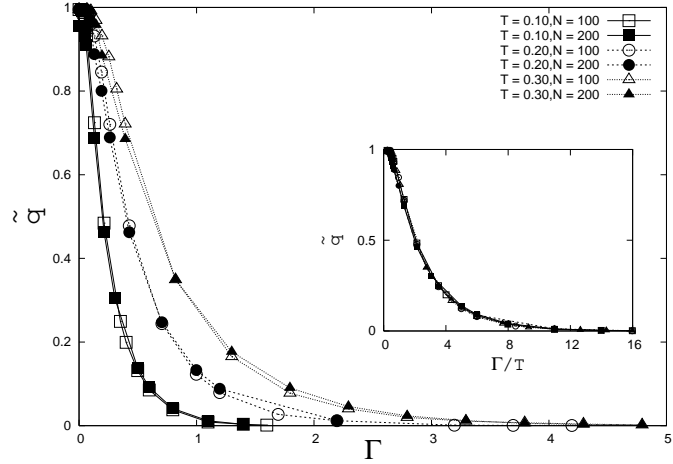}
\caption{\footnotesize
\label{fig:insetq}Variation of the order parameter $\tilde{q}$
(correlation in the Trotter direction)
with transverse field $\Gamma$ for $T = 0.10, 0.20$ and
$0.30$ ($h = 0$) for two different system sizes ($N = 100$ and $200$).
$\tilde{q} = 0$ for large  $\Gamma $.
 The inset shows the plot of $\tilde{q}$ against the scaled variable $\Gamma/T$.}
\ec
\end{figure}
\begin{figure}
\bc
\noindent \includegraphics[clip,width= 9cm, angle = 0]
{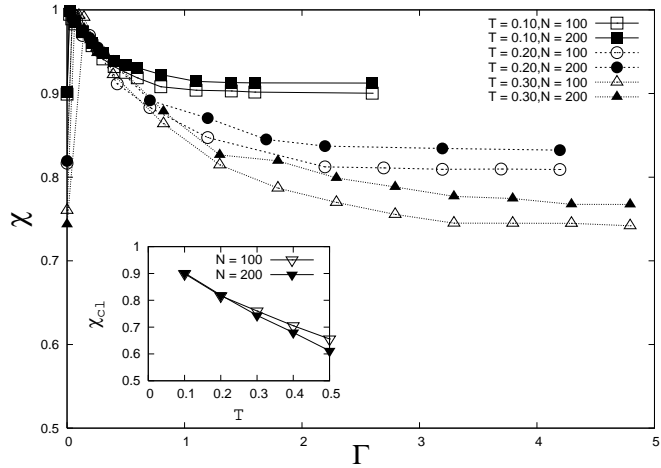}
\caption{\footnotesize
\label{fig:inset5}Variation of the
susceptibility $\chi$ with transverse field $\Gamma$ for
$T = 0.10, 0.20$ and $0.30$ ($h \le 0.1$) for two different system sizes
($N = 100$ and $200$). The corresponding susceptibility $\chi_{cl}$ for various
temperatures for $N = 100$ and $200$ for the
classical system are shown in the inset. $\chi$
converges to the classical values $\chi_{cl}$ for
large $\Gamma$.}
\ec
\end{figure}
\begin{figure}
\bc
\noindent \includegraphics[clip,width= 9cm, angle = 0]{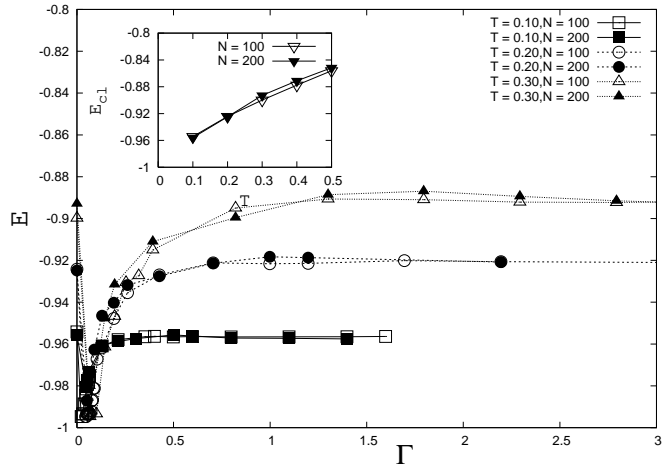}
\caption{\footnotesize
\label{fig:inset4}Variation of average energy $E$ with
transverse field $\Gamma$ for $T = 0.10, 0.20$ and $0.30$ ($h = 0$) for
two different values
of $N$($=100,200$). The corresponding average energy $E_{cl}$ for various
temperatures for $N = 100$ and $200$ for the are shown in the inset.
$E$ converges to the classical values $E_{cl}$ for large $\Gamma$ .  }
\ec
\end{figure}
\subsubsection*{\bf II.A.2. Monte Carlo simulation}
\noindent This Hamiltonian (\ref{eq:Hamiltonian1}) can be mapped to a
$(\infty + 1)$-dimensional classical
Hamiltonian \cite{Suzuki,RMP-Das}
using the Suzuki-Trotter formula. The effective Hamiltonian
can be written as (for $J_0 = -1$)
\begin{eqnarray}
{\mathcal H}  & = & - \frac{1}{NM}  \sum_{i,j(>i)=1}^N 
\sum_{k=1}^M \sigma_{i,k} \sigma_{j,k} - \frac{h}{M} 
\sum_{i=1}^N \sum_{k=1}^M \sigma_{i,k} \notag \\ & + & B\sum_{i=1}^N 
\sum_{k=1}^M \sigma_{i,k} \sigma_{i,k+1},
\end{eqnarray}
where
\begin{equation}
B =  (1/2) \ln(\coth (\Gamma/MT)) .
\end{equation}
Here $M$ is the number of Trotter replicas and $k$ denotes the
$k$-th row in
the Trotter direction. $B$ denotes the nearest-neighbor interaction
strength along the Trotter direction. We have studied the system for $N = 100$.
Because of the diverging growth of interaction $B$ for very low values of
$\Gamma$ and also for high values of $M$, and the consequent non-ergodicity
(the system relaxes to different states for identical thermal and quantum parameters, due to frustrations, starting from
different initial configurations), we have kept
the value of $M$ at a fixed value of $5$.
This choice of $M$ value helped satisfying the ergodicity
of the system up to
very low values of the transverse field at the different
temperatures considered
$T = 0.10, 0.20$ and $0.30$.
Starting from a random initial configuration
(including all up or 50-50
up-down configurations) we
follow the time variations of different
quantities until they relax and
study the various quantities after they relax.\\
We studied results for three different temperatures
$T = 0.10, 0.20$ and $0.30$ and all the results
are for $N = 100$ and $200$ and $M = 5$.
We estimated the following quantities after relaxation :

\medskip
(i) {\em Correlation along Trotter direction ($\tilde{q}$)} :
We studied the variation
of the order parameter
\begin{equation}
\tilde{q} = \frac {1}{NM}\sum_{i=1}^N \sum_{k=1}^M \langle \sigma_{i,k} \sigma_{i,k+1}  \rangle,
\end{equation}
which is the first neighbor correlation along Trotter direction. Here,
$\langle ... \rangle$ indicate the average over initial
spin configurations. This quantity $\tilde{q}$ shows a smooth vanishing 
behavior.
We consider this correlation $\tilde{q}$ as the order parameter for the 
transition at
$\Gamma_c$. A larger transverse field is needed for the vanishing of the
order parameter for larger temperature. The observed values
(see Fig. \ref{fig:insetq}) of $\Gamma_c$
are $\simeq 1.6, 2.2$ and $3.0$ for $T = 0.1, 0.2$ and $0.3$ respectively.
As shown in the inset, an unique data collapse occurs when $\tilde{q}$ is plotted
against $\Gamma/T$ and one seems to get the complete disorder immediately as
the scaling dos not involve any finite value $T_c$. This is consistent
with the observations in the next section.

\medskip

(ii) {\em Susceptibility ($\chi$)} : The longitudinal
susceptibility $\chi = (1/NM) {\partial [\sum_{i,k} 
\langle\sigma_{i,k}\rangle}]/{\partial h}$,
where $h$ ($\rightarrow 0$) is the applied longitudinal field, has also been
measured. We went up to $h = 0.1$ and estimated the $\chi$ values. As we
increase the value of the
transverse field $\Gamma$ from a suitably chosen low value, $\chi$ initially
starts with a value almost equal to unity and then gradually saturates at lower
values (corresponding to the classical system where $B = 0$ in Eq.(14)) as
$\Gamma$ is increased. Also at $\Gamma = 0$, the classical values are
indicated in Fig. \ref{fig:inset5}. This
saturation value of $\chi$ decreases with temperature. Again the field at
which the susceptibility saturates are the same as for the vanishing of the
order parameter for each temperature.
\medskip

(iii) {\em Average energy (E)} : We have measured the value of
the
co-operative energy for each Trotter index and then take its average $E$ i.e.
$E = \langle H^{(C)} \rangle$ of Eq. (10) with $J_0 = -1$. It initially begins
with $-1.0$ and after a sharp rise the average energy saturates,
at large values of $\Gamma$, to
values corresponding to the classical equilibrium energy ($E_{cl}$ for 
$B = 0$ in Eq.(14)) at those
temperatures. Again it takes larger values of $\Gamma$ at higher temperatures
to achieve the classical equilibrium energy. At $\Gamma = 0$, the corresponding
classical values of $E$ are plotted in Fig. \ref{fig:inset4}.
The variations of all these quantities indicate that the `quantum order'
disappears
and the quantities reduce to their classical values
(corresponding to $B = 0$) for large values of
the transverse field $\Gamma$.

The continuous transition-like behaviour seen from Fig. \ref{fig:insetq} can
be justified from a mean field analysis (see Appendix B). At finite temperature
it is the free energy that we have to minimise and the entropy term plays a 
crucial role. Minimization of free energy leads to an analytic variation of
the total magnetization and no phase transition at any finite temperature.

\subsection*{\bf II.B. LRIAF with disorder : Classical model at finite 
temperature}
\subsubsection*{\bf II.B.1. Analytical studies}
\noindent  For the classical case, i.e., $\Gamma = 0$, Eqns. 
(\ref{eq:quantum_mz}) and (\ref{eq:quantum_q}) reduce 
to (for $h=0$)\\
\begin{eqnarray}
m & = & \int_{-\infty}^{\infty}Dy \tanh \beta (\tilde{J}\sqrt{q}y +J_{0}m) \\
q & = &  
\int_{-\infty}^{\infty}
Dy \tanh^2 (\beta \tilde{J}\sqrt{q}y + J_{0}m) ~~,
\end{eqnarray}
where $m^z$ has been replaced by $m$.
For $J_{0} < 0$, again we find that $m=0$ is only physical solution
for all temperature regimes.
This means that there are three possible
phases : namely, the antiferromagnetic phase, the paramagnetic phase and
the spin glass phase. In all these three phases, the magnetization $m$ is zero.
To determine the critical point $T_{SG}$ at which the spin glass transition
takes place, we expand the equation with respect to $q$ for $q \simeq 0$ and 
$m=0$. 
In the limit of $\tilde{J} \ll 1$, we have
\be
q  =  
\frac{(\beta \tilde{J})^{2}-1}
{2(\beta \tilde{J})^{4}} ~~.
\ee
\noindent We therefore have $T_{SG} = \tilde{J}$ and the critical point is 
independent
of the antiferromagnetic bias $J_{0}$.
\begin{figure}[ht]
\begin{center}
\includegraphics[width=9cm]{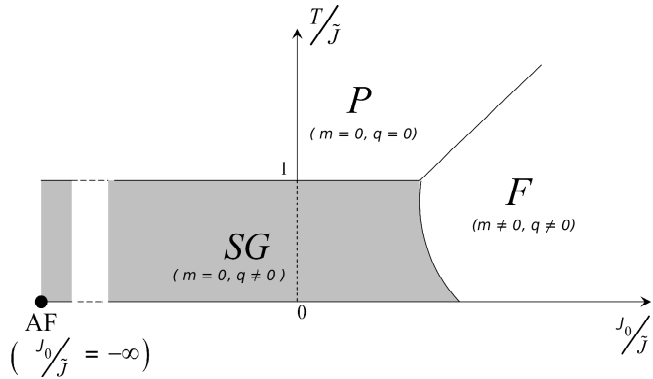}
\end{center}
\caption{\footnotesize
The phase diagram of classical SK model \cite{Binder} extended for 
antiferromagnetic bias. For $J_{0}<0$, there exist spin glass phase
below $T/\tilde{J}=1$ and the critical temperature is independent of the 
strength of the antiferromagnetic bias $J_{0}$. For pure LRIAF at finite 
temperature ($T>0$),
the anti-ferromagnetic order disappears and the system changes to the 
paramagnetic phase. When we add an infinitesimal disorder $\tilde{J}>0$,
the antiferromagnetic order is broken down and the system suddenly gets 
`frozen' into a spin glass (SG) phase.}
\label{fig:fg2}
\end{figure}
\mbox{}
This result means that the antiferromagnetic order can appear if and only if
we set $J_{0}<0$ and $T/\tilde{J}=0,J_{0}/\tilde{J}=-\infty$.
On the other hand, for $-\infty < J_{0} < 0$ at low temperature regime
$T < T_{SG}$, the spin glass phase appears.
We plot the phase diagram in Fig. \ref{fig:fg2}.
We also conclude that the system described by the Hamiltonian 
(\ref{eq:Hamiltonian1}) with $\Gamma =0$ is immediately `frozen' when
we add any infinitesimal disorder $\tilde{J} >0$.

From the view point of the degeneracy of the spin configurations,
we easily estimate the number of solution for the antiferromagnetic
phase as $N!/[(N/2)!(N/2)!] \simeq {\rm e}^{0.693N}$ (in the antiferromagnetic 
ground 
state only $N/2$ spins out of $N$ will have free choice (to be up or down) and 
the rest have to follow), which is 
larger than the number of the SK model ${\rm e}^{0.199N}$ \cite{Binder}.
However, for the infinite range antiferromagnetic model, the energy barrier 
between arbitrary configurations which gives the same
lowest energy states is of order $1$ and there is no ergodicity breaking.

\subsubsection*{\bf II.B.2 Monte Carlo studies}
\noindent In order to study the effect of introducing disorder in the 
classical LRIAF
model, we performed a Monte Carlo study with a system of $N=100$ spins. But 
the distribution function ($\tilde{P}$) of disorder introduced here is 
different. Instead of a 
Gaussian distribution, we applied a binary distribution ($\pm J$) with a 
probability $p$ :
\be
\tilde{P}(J_{ij}) = p \delta (J_{ij} - J) + (1 - p) \delta (J_{ij} + J) ~~.
\ee
In our study we kept $J = 1$ and $J_0 = \bar{J}$, where $\bar{J}$ is the 
average interaction strength. Each of the Ising spins interact with every other
ferromagnetically with probability $p$ and antiferromagnetically with 
probability $1-p$. Therefore $\bar{J} = 2p - 1$ and 
$\tilde{J} = 2\sqrt{p(1-p)}$. At the two limiting values
of $p = 0$ and $p = 1$, the system becomes purely antiferromagnetic (LRIAF) and 
purely ferromagnetic respectively. Thus at these two limiting values of $p$, 
the system has no fluctuation at all ($\tilde{J} = 0$). Whereas for $p = 0.5$, 
the fluctuation is maximum ($\tilde{J} = 1$). So as predicted above, we need 
maximum thermal fluctuation to destroy the glassy phase.

\begin{figure}[h]
\begin{center}
\includegraphics[width=9cm]{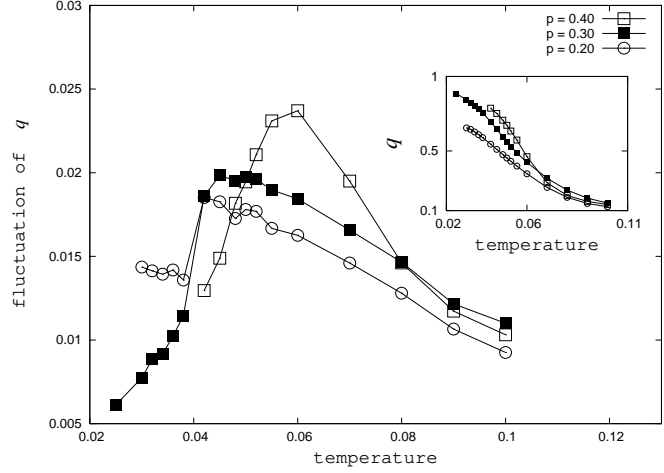}
\end{center}
\caption{\footnotesize\label{fig:insetcorr}
Variation of the fluctuation of spin-spin correlation $q$ 
with temperature $T$ for $p = 0.40, 0.30$ and $0.20$ for $N = 100$ for 
classical disordered system. The inset
shows the plot of $q$ against temperature $T$.
}
\end{figure}

To identify the glass phase, we considered a replica of the original system to 
be studied and evolved the two systems simultaneously by Monte Carlo technique.
We also repeated the study for $N = 50$ and $200$ also. For all sizes we 
obtained almost same phase diagram. We measured the average
absolute value of the spin-spin correlations of the two systems (the original
and the replica one) at 
different times for a certain temperature. Let us denote this parameter by 
$q$ and $\alpha$ and $\beta$ denotes the original system and
the replica.
\be
q = \left\langle \left|\frac{1}{N}\sum_{i=1}^N s_i^\alpha s_i^\beta \right| \right\rangle ~~.
\ee
We measure the average steady state values of this parameter for various 
temperatures. The average is over different configurations. When $p$ is 
very high i.e., when the number of ferromagnetic bonds are dominant, the 
ferromagnetic to
paramagnetic transition can be identified easily by measuring the change in
magnetisation and divergence of susceptibility at the critical temperature.
But for $p < 0.5$ and slightly above $0.5$, the magnetisation remains low for
all temperatures and hence it is difficult to identify the existence of any
glass phase/paramagnetic phase seperately. The glass and paramagnetic phase
can be distinguished by studying the steady state values and fluctuation of 
$q$. We have studied for three different values of $p = 0.20,
0.30$ and $0.40$. For a particular value of $p$, upto a certain 
temperature the value of $q$ remains high indicating the
spin glass phase. The fluctuations in $q$ remain small.
This implies that the flipping of spins is very low (freezing of spins).
As the temperature is raised, the steady state value of $q$
decreases gradually (inset of Fig. \ref{fig:insetcorr}) but the fluctuation 
increases. At a certain temperature
(glass-para phase transition point) this fluctuation reaches maximum and
above this temperature the value of $q$ gradually goes to 
zero. With further increase of temperature, the fluctuation decreases 
indicating a second order glass-para phase transition 
(Fig. \ref{fig:insetcorr}).

\begin{figure}[h]
\begin{center}
\includegraphics[clip,width=7cm, angle = 270]{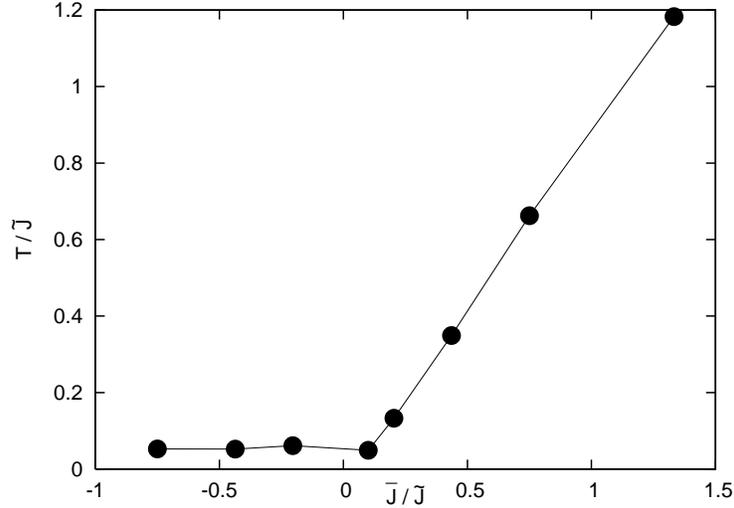}
\end{center}
\caption{\footnotesize\label{fig:phase.l100}
The phase diagram of classical LRIAF with disorder obtained from Monte Carlo
simulation. This phase diagram is similar to that obtained from static and
replica symmetric approximations. 
}
\end{figure}

It can be observed from (Fig. \ref{fig:insetcorr}), that as we increase the 
value of $p$ from $0$ to $0.5$,
the transition temperature increases. It is expected from our previous 
explanation, as $p$ approaches $0.5$, fluctuations $\tilde{J}$ increases and 
so is the transition temperature. We have given a plot of the $T_c/\tilde{J}$
as a function of $\bar{J}/\tilde{J}$ (Fig. \ref{fig:phase.l100}). 

\subsection*{\bf II.C. LRTIAF with disorder : Quantum model at finite temperature}

\subsubsection*{\bf II.C.1. Analytical studies}

\begin{figure}[ht]
\begin{center}
\includegraphics[clip,width=6cm,angle=270]{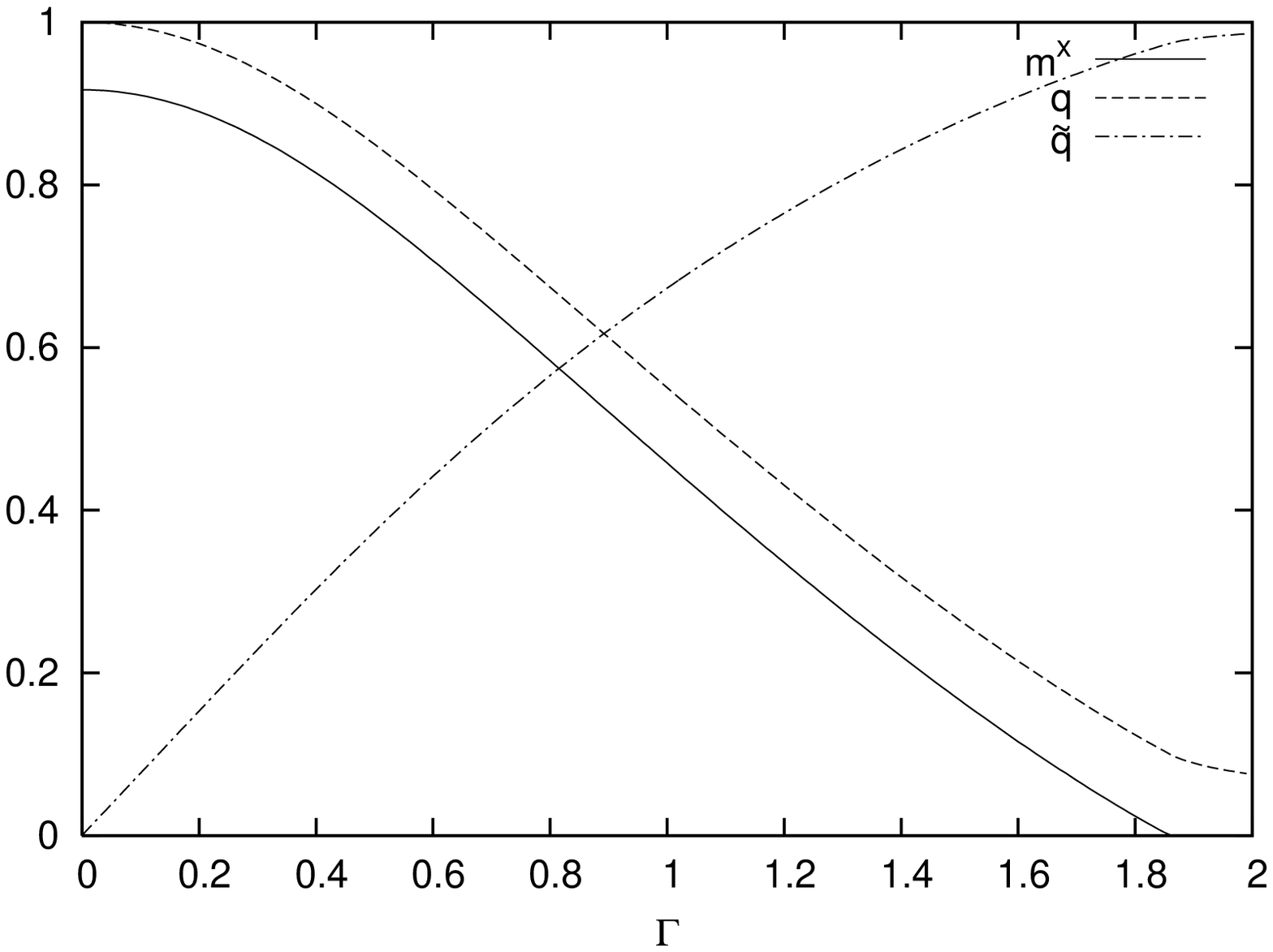}
\end{center}
\caption{\footnotesize\label{fig:beta10.dat}
The result of numerical calculations for
the saddle point equations for
$m_{x}$, $q$ and $\tilde{q}$ as a function of
$\Gamma$ for $T\ne0$.
}
\end{figure}
\begin{figure}[ht]
\begin{center}
\includegraphics[width=8cm]{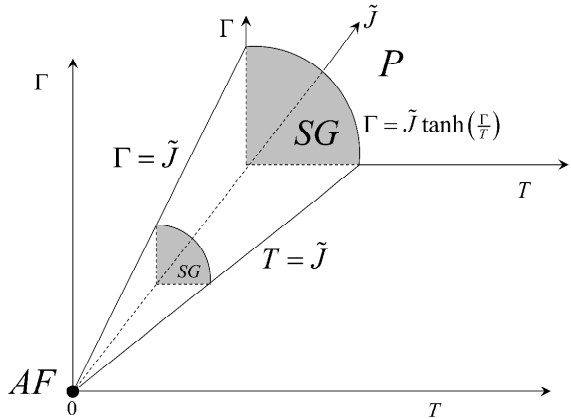}
\end{center}
\caption{\footnotesize\label{fig:3dphased2}
Phase diagram for the quantum system. The antiferromagnetic order exists if 
and only if we set $T=\Gamma=0$ and $\tilde{J} = 0$. As the $\tilde{J}$ 
decreases, the spin glass 
phase gradually shrinks to zero and eventually ends up at an antiferromagnetic 
phase at its vertex (for $\Gamma=0=T=\tilde{J}$) as discussed in 
Section II.C.1.
}
\end{figure}

\noindent The approximate saddle point equations have already been presented in 
equations \ref{eq:quantum_mz}, \ref{eq:quantum_mx} and \ref{eq:quantum_q}.
For detail calculations see Appendix A.
The variations of $m^x$, $q$ and $\tilde{q}$ are shown in Fig.
\ref{fig:beta10.dat}.
The phase boundary between
the spin glass and paramagnetic phases is given by setting $m^z=0$ and 
$q \simeq 0$ and we get \\
\begin{equation}
\Gamma = \tilde{ J}\tanh \left ( \Gamma \over T\right)
\label{eq:boundary}.
\end{equation}
Obviously, the boundary at $T=0$ gives $\Gamma_{SG}=\tilde{J}$.
On the other hand, when we consider the case of $\Gamma \simeq 0$, we have
$T_{SG}=\tilde{J}$ (consistent with the classical result). These facts imply 
that there is neither an antiferromagnetic nor
a spin glass phase when we consider the pure case $\tilde{J}=0$ because
the critical point leads to $T_{SG}=\Gamma_{SG}=0$. Therefore,
we conclude that the antiferromagnetic phase can exist if and only if
$T=\Gamma=0$ (Fig. \ref{fig:3dphased2}).
\subsubsection*{\bf II.C.2. Quantum Monte Carlo studies}
\noindent Now to study the quantum system we again apply the finite 
temperature quantum 
Monte Carlo method as applied to study the pure LRIAF model in Section II A.
Like the classical system, here also we incorporated a disorder with binary
distribution ($\pm J$) with a probability $p$. 
As mentioned earlier, to study by quantum Monte Carlo, we map the Hamiltonian
(\ref{eq:Hamiltonian}) to a ($\infty + 1$) dimensional classical Hamiltonian
using the Suzuki-Trotter formula. The effective Hamiltonian can be written
as (for $h=0$)\\
\begin{equation}
{\mathcal H} = - \frac{1}{NM} \sum_{i,j(>i)=1}^N
\sum_{k=1}^M J_{ij} \sigma_{i,k} \sigma_{j,k} +B  \sum_{i=1}^N
\sum_{k=1}^M \sigma_{i,k} \sigma_{i,k+1},
\end{equation}
where
\begin{equation}
B = (1/2)\ln(\coth (\Gamma/MT)) .
\end{equation}
$M$, $k$ and $B$ represents the number of Trotter replicas, $k$-th row in
Trotter direction and nearest-neighbour interaction strength along the
Trotter direction respectively. The distribution $\tilde{P}$($J_{ij}$) is
given by Eq. (20).
To identify the glass phase, a replica of the original system to be studied
has been taken and the two systems has been evolved simultaneously by Monte
Carlo technique.

\begin{figure}[h]
\begin{center}
\subfigure{
\includegraphics[width=8cm]{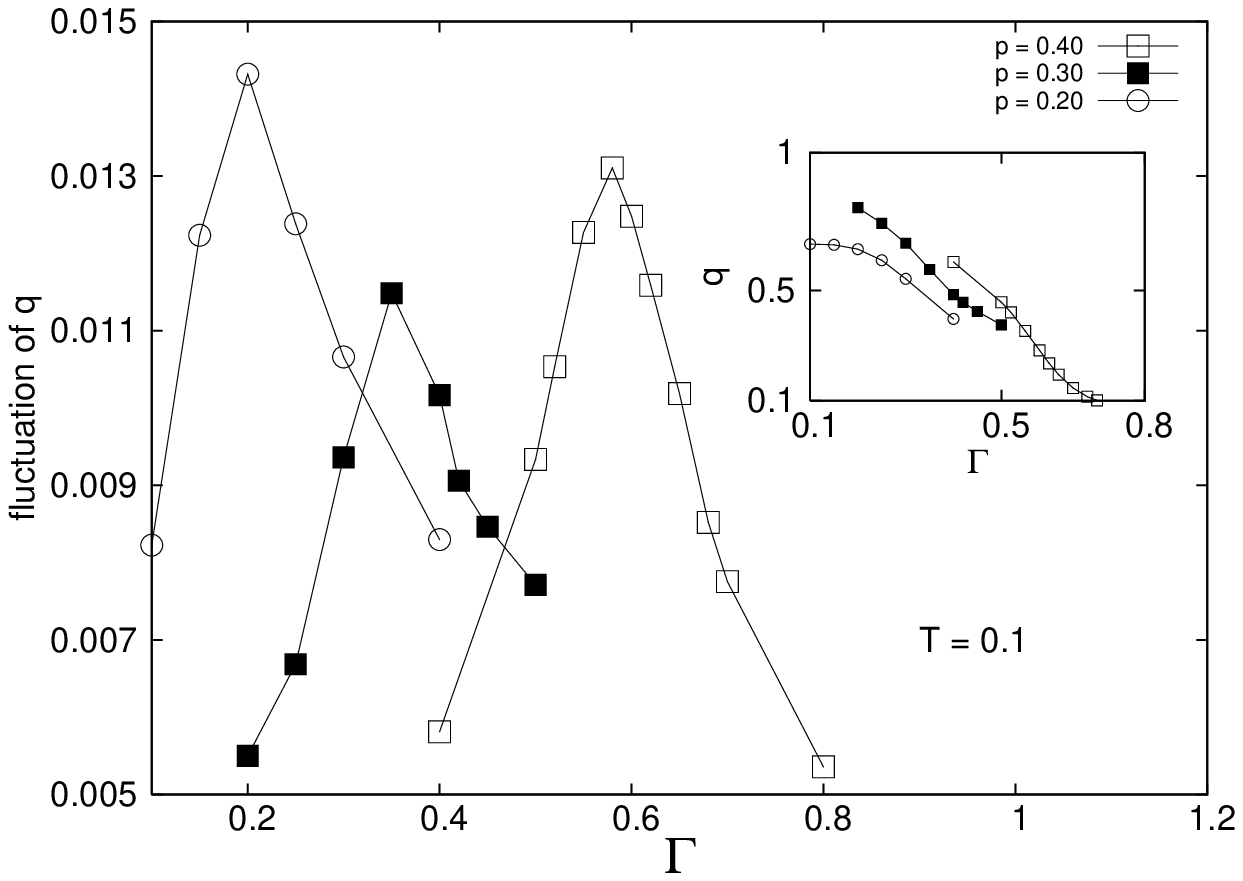}\label{fig:insetcorrq}}
\subfigure 
{\includegraphics[width=8cm]{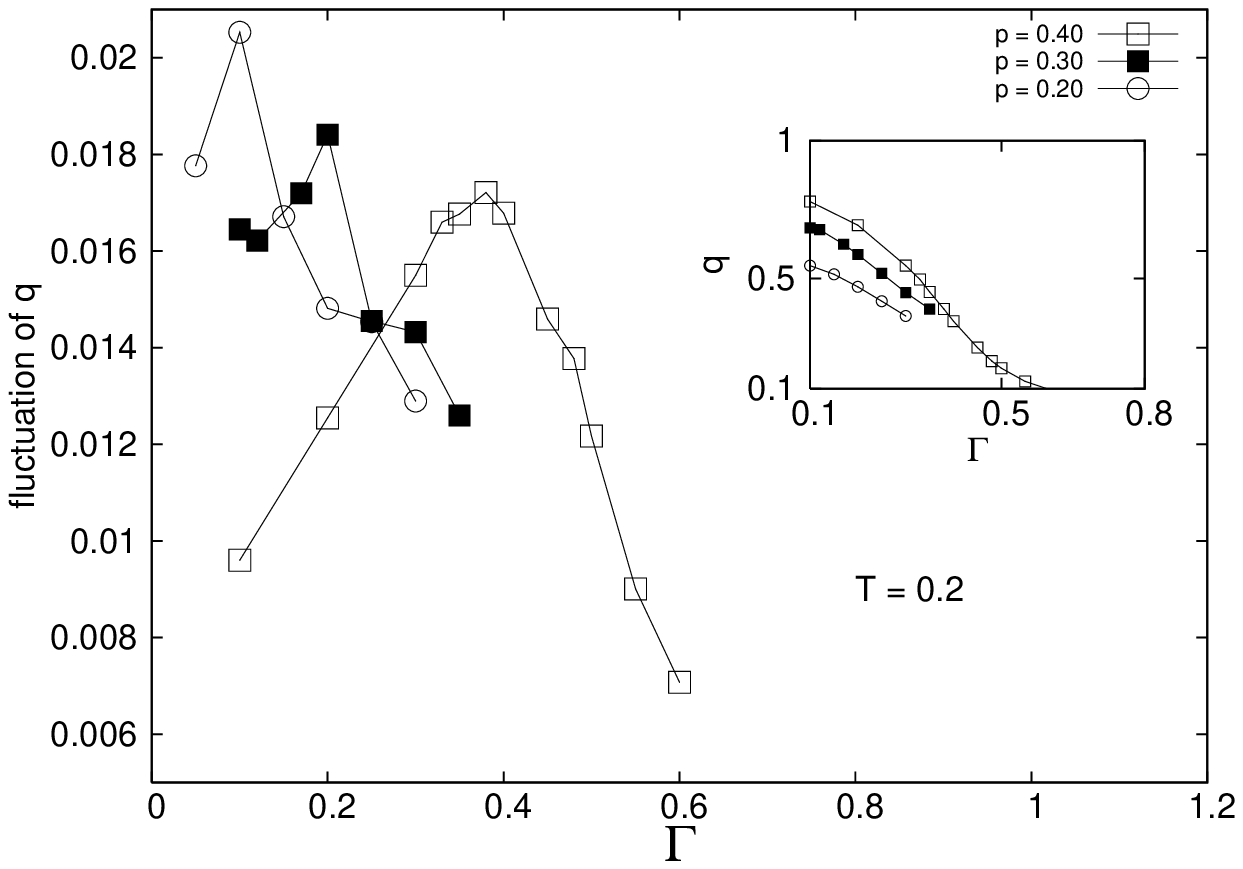}\label{fig:insetcorrq.T0.2}}
\label{fig:subfigureExample}
\caption[Optional caption for list of figures]{\subref{fig:insetcorrq}Variation of the fluctuation of spin-spin correlation
$q$
with $\Gamma$ at $T=0.1$ for $p = 0.40, 0.30$ and $0.20$ for $N = 100$ 
($M = 5$). The
inset shows the plot of $q$ against transverse field
$\Gamma$. \subref{fig:insetcorrq.T0.2} Variation of the fluctuation of 
spin-spin correlation $q$
with $\Gamma$ at $T=0.2$ for $p = 0.40, 0.30$ and $0.20$ for $N = 100$. The
inset shows the plot of $q$ against transverse field $\Gamma$.}
\end{center}
\end{figure}
\begin{figure}[h]
\begin{center}
\includegraphics[clip,width=8cm,angle=0]{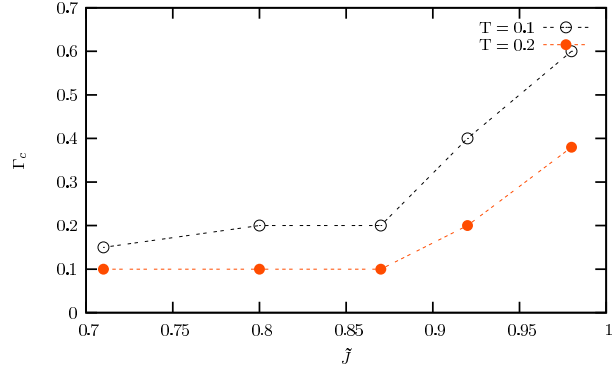}
\end{center}
\caption{\footnotesize\label{fig:qphase}
Variation of $\Gamma_c$ with $\tilde{J}$ for the quantum LRIAF with disorder
obtained from Monte Carlo simulation. This plot is qualitatively similar to 
that obtained from static and replica symmetric approximations.
}
\end{figure}

The quantity measured ($q$) is the average
absolute value of the spin-spin correlations of the two replica systems at
different times for a certain temperature and $\Gamma$, where $\alpha$ and
$\beta$ denotes the original one and the replica.
\be
q = \left\langle \left|\frac{1}{NM}\sum_{i=1}^{NM} s_i^\alpha s_i^\beta \right| \right\rangle ~~.
\ee
Initially we kept the temperature fixed at a certain value (typically $T=0.1$)
and measured the average steady state values of this parameter for various
values of $\Gamma$. The average is over different configurations. Here also 
the glass and paramagnetic phase
can be distinguished by studying the steady state values and fluctuation of
$q$. For a particular value of $p$, upto a certain
$\Gamma$ the value of $q$ remains high indicating the
spin glass phase. The fluctuations in $q$ remain small.
This implies that the flipping of spins is very low (freezing of spins).
As the value of $\Gamma$ is raised, the steady state value of
$q$
decreases gradually (inset of Fig. \ref{fig:insetcorrq}), but the 
fluctuation increases (Fig. \ref{fig:insetcorrq}). At a certain value of
$\Gamma$ (say $\Gamma_c$)
(glass-para phase transition point) this fluctuation reaches maximum and
above this $\Gamma_c$ the value of $q$ goes to zero. With
further increase of $\Gamma$, the fluctuation decreases indicating a
second order glass-para phase transition.
The observed values of $\Gamma_c$ are $\simeq 0.2, 0.35$ and $0.58$ for 
$p = 0.2, 0.3$ and $0.4$ respectively. A similar study for $T = 0.2$ is
presented in Fig. \ref{fig:insetcorrq.T0.2}. The values of $\Gamma_c$ are 
$\simeq 0.10, 0.20$ and $0.40$ for $p = 0.2, 0.3$ and $0.38$ respectively.

It has been noticed, that as we increase the value of $p$ from $0$ to $0.5$,
the transition field increases. It is expected from our previous
explanation, as $p$ approaches $0.5$, fluctuations $\tilde{J}$ increases and
so is the transition field $\Gamma_c$. We have given a plot of the $\Gamma_c$
as a function of $\tilde{J}$ (Fig. \ref{fig:qphase}). Though these results 
do not satisfy the predictions of static and replica symmetric approximations
quantitatively, but qualitatively they reflect all the features.

\subsection*{\bf IID. LRTIAF with disorder : Quantum model at zero temperature}
\noindent 
We study this case only analytically.
As well-known, in the mean-field description of 
the pure quantum transverse Ising systems, 
the total magnetization is conserved as 
$(m^{z})^{2}+(m^{x})^{2}=1$. 
However, if some disorders are taken into account, 
it is non-trivial problem to answer the question; if the magnetization conservation 
still holds or not. As we discussed before, for antiferronmagnets, 
$m_{z}$ is always zero and the magnetization conservation reads $(m^{x})^{2}=1$. 
In following, we derive the condition 
on which the magnetization conservation holds. 

\begin{figure}[ht]
\begin{center}
\includegraphics[clip,width=6cm,angle=270]{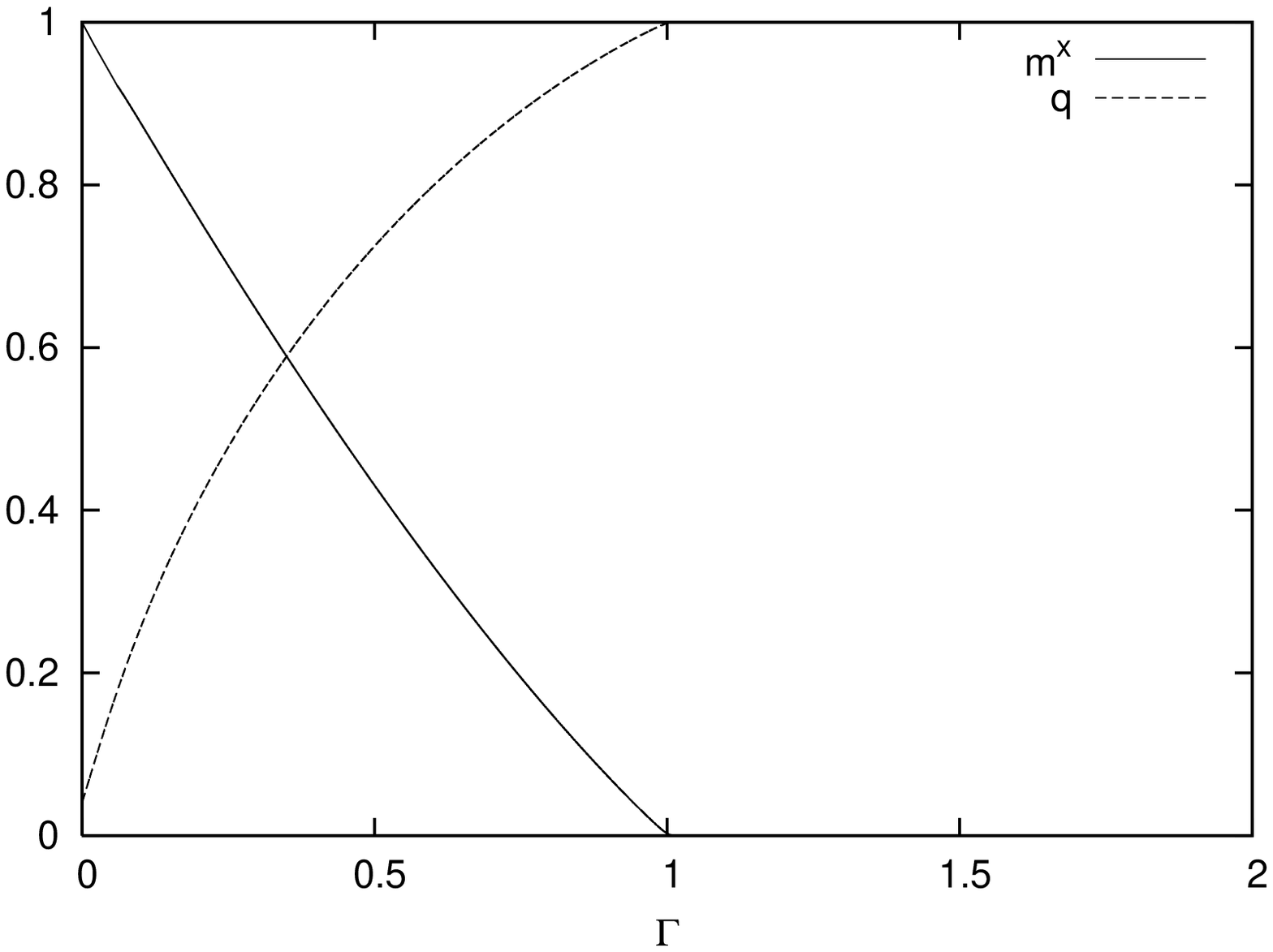}
\end{center}
\caption{\footnotesize\label{fig:zero.dat}
The result of numerical calculations for
the saddle point equations for
$m_{x}$ and $q$ as a function of
$\Gamma$ for $T=0$.
}
\end{figure}

For finite transverse field but zero temperature, i.e., $\beta = \infty$,
equations (8) and (9) reduce to 
\begin{eqnarray}
m^{x} & = &
\int_{-\infty}^{\infty}
Dy
\frac{\Gamma}
{\sqrt{(\tilde{J}\sqrt{q}y)^{2}+
\Gamma^{2}}} \\
q & = &
\int_{-\infty}^{\infty}
Dy
\left\{
\frac{\tilde{J}\sqrt{q}y}
{\sqrt{(\tilde{J}\sqrt{q}y)^{2}+
\Gamma^{2}}}
\right\}^{2}  
\label{eq:equation_q}
\end{eqnarray}
which are obtained within the replica symmetric theory 
in this zero temperature limit (see Appendix A).
The variations of $m^x$ and $q$ are shown in Fig.
\ref{fig:zero.dat}.
It can also be seen that, for no disorder, i.e. $\tilde{J} = 0$, we obtain
$m^x = 1$ and $q = 0$.  
For finite disorder $\tilde{J} \neq 0$, 
we expand equation (\ref{eq:equation_q}) with respect to $\tilde{J}\sqrt{q}y  \ll 1$. 
Then, we have $1\simeq (\tilde{J} / \Gamma)^{2}-3q(\tilde{J} / \Gamma)^{4}$, 
namely, 
\begin{eqnarray}
q & = & 
\frac{
\left(
\frac{\tilde{J}}{\Gamma}
\right)^{2}-1}
{
3\left(
\frac{\tilde{J}}{\Gamma}
\right)^{4}
},
\label{eq:resultQ}
\end{eqnarray}
where we used $\int_{-\infty}^{\infty}
Dy y^{2} =  1,\, \int_{-\infty}^{\infty}
Dy y^{4}=3$. 
Eqn. (\ref{eq:resultQ}) implies $q \ne 0$ for $\Gamma < \tilde{J}$.
Hence,
\begin{eqnarray}
m^{x} & = &
1-\frac{1}{6\tilde{J}}
\left(\frac{\left({\tilde{J}}/{\Gamma}\right)^{2}-1}{{\tilde{J}}/{\Gamma}}
\right) ,\\
q & = & 
\frac{\left({\tilde{J}}/{\Gamma}\right)^{2}-1}{3\left({\tilde{J}}/{\Gamma}\right)^{4}} 
\end{eqnarray}
\begin{figure}[ht]
\begin{center}
\includegraphics[clip,width=6cm,angle=270]{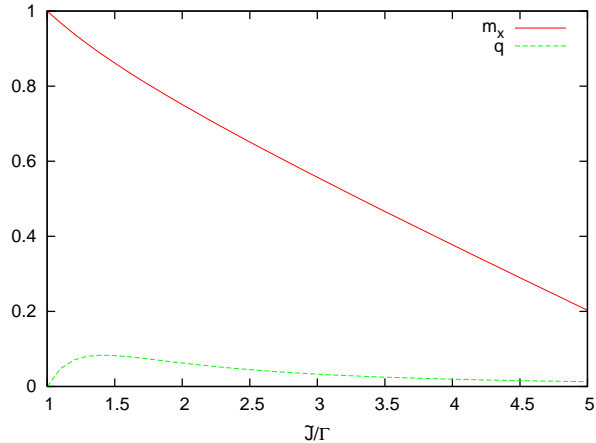}
\end{center}
\caption{Variation of $m^x$ and $q$ with $\tilde{J}/{\Gamma}$}
\label{fig:fg3}
\end{figure} 
From this result, we find that 
magnetization conservation 
is broken for $\Gamma < \tilde{J}$. 
This fact means that 
each spin starts `frozen' locally and the degree of 
freedom of spins is remarkably reduced. 
Therefore, $m^{x}$ might be a good indicator to 
detect the spin glass phase for the system 
in which the both order (antiferromagnet) and disorder 
(spin glass) phases possess the same spatial symmetry  with $m^{z}=0$. 
\section{Discussion}

\noindent We considered here first a long-range Ising antiferromagnet with 
disorder incorporated in it and put it in a transverse field. Although 
long-range interaction is unrealistic and also difficult for numerical
studies, it is convenient for analytical approaches such as mean-field 
calculations which are exact in certain limits. So here we have studied
this kind of long-range interaction. 
We have obtained the finte temperature
free-energy expression (Appendix A) for this model and studied analytically 
the magnetisation, spin glass order and the 
correlation (Trotter replica). 
For the pure case (i.e. no disorder) the antiferromagnetic
order is seen to get immediately broken as soon as the thermal or
quantum fluctuations are added (see Sec. II.A.).
However, when we add the disorder as in SK Hamiltonian, on that of LRTIAF as
perturbation, we find that an infinitesimal spin glass disorder is enough to
induce a stable glass order in this LRTIAF antiferromagnet (Sec. II.B.). This 
glass order eventually gets destroyed as
the thermal or quantum fluctuations increased beyond their threshold values and
the transition to para phase occurs (Sec. II.B. and II.C.).
As shown in the phase diagram in Fig. \ref{fig:fg2},  the antiferromagnetic 
phase of the 
LRTIAF (occurring only at $\tilde{J} = 0 = \Gamma = T$), can get `frozen' 
into spin-glass phase if a little SK-type disorder is
added ($\tilde{J} \neq 0$); the only missing element in the LRTIAF
(which is fully frustrated, but lacks disorder)
to induce stable order (freezing of random spin orientations) in it.
These results has been confirmed by Monte Carlo simulations.
We have not addressed the question of possible replica symmetry breaking in 
our study.
The degeneracy factor ${\rm e}^{0.693N}$
of the ground state of the LRIAF is much larger than that ${\rm e}^{0.199N}$ 
for the SK model. Hence, (because of the presence of
full frustration) the LRIAF possesses a surrogate incubation property of
stable spin glass phase in it when induced by addition of a small disorder.

\begin{acknowledgements}
We are grateful to I. Bose, A. Das, S. Dasgupta, D. Sen,
P. Sen and K. Sengupta for useful discussions and comments. 
One of the authors (JI)  were financially supported by Grant-in-Aid Scientific Research on
Priority Areas {\it `Deepening and Expansion of Statistical Mechanical Informatics
(DEX-SMI)'} of the MEXT No. 18079001 and  
INSA (Indian National Science Academy) -  JSPS 
(Japan Society of Promotion of Science)  Bilateral Exchange Programme. 
He also thanks Saha Institute of Nuclear Physics for their warm hospitality during 
his stay in India. 
AKC acknowledges Hokkaido University for their hospitality during his stay in Japan.

\end{acknowledgements}

\appendix

\section{Derivation of free energy}
In this appendix, we show the derivation 
of the free energy per spin for the system to be described by 
the Hamiltonian.
\begin{eqnarray}
H & = & 
-\sum_{ij}J_{ij}\sigma_{i}^{z}\sigma_{j}^{z} - 
\Gamma \sum_{i}\sigma_{i}^{x}. 
\end{eqnarray}
Carrying out the Suzuki-Trotter decomposition, we have 
the replicated partition function. 
\begin{eqnarray}
Z_{M}^{n} & = & 
{\rm tr}_{\{\sigma\}}
{\exp}
\left[
\frac{\beta}{M}
\sum_{ij}
\sum_{k}
\sum_{\alpha}
J_{ij} \sigma_{i}^{\alpha}(k) 
\sigma_{j}^{\alpha} (k) + 
B \sum_{i}\sum_{k}\sum_{\alpha}
\sigma_{i}^{\alpha} (k) \sigma_{i}^{\alpha} (k+1)
\right] \\
B & = & 
\frac{1}{2}
\ln \coth 
\left(
\frac{\beta \Gamma}{M}
\right)
\end{eqnarray}
where 
$\alpha$ and $k$ denote the replica and Trotter indices. 
$M$ is the number of the Trotter slices and 
$\beta$ is the inverse temperature. 
The disorder $J_{ij}$ obeys
\begin{eqnarray}
P(J_{ij}) & = & 
\frac{1}{\sqrt{2\pi J^{2}}}
{\exp}
\left[
-\frac{(J_{ij}-j_{0})^{2}}
{2J}
\right]
\end{eqnarray}
In other words, the $J_{ij}$ follows 
\begin{eqnarray}
J_{ij} & = & j_{0}+ J x, \,\,\,\,P(x) = \frac{1}{\sqrt{2\pi}}{\rm e}^{-\frac{x^{2}}{2}}.
\end{eqnarray}
We should notice that 
$j_{0} >0, J=0$ is 
pure ferromagnetic transverse Ising model, whereas 
$j_{0}<0, J=0$ corresponds to
pure antiferromagnetic transverse Ising model.   
Then,  by using 
$\int_{-\infty}^{\infty}Dx\,{\rm e}^{ax}={\rm e}^{a^{2}/2}$, 
$Dx \equiv dx \,{\rm e}^{-x^{2}/2}/\sqrt{2\pi}$, 
we have the average of the replicated partition function as 
\begin{eqnarray}
\ll 
Z_{M}^{n} \gg & = & 
{\rm tr}_{\{\sigma\}}
{\exp}
\left[
\frac{\beta j_{0}}{M}
\sum_{k}
\sum_{\alpha}
\sum_{ij}
\sigma_{i}^{\alpha}(k)
\sigma_{j}^{\alpha}(k)
+ 
B\sum_{k}\sum_{\alpha}\sum_{i}
\sigma_{i}^{\alpha}(k)
\sigma_{i}^{\alpha}(k+1) 
\right] \nonumber \\
\mbox{} & \times & 
{\exp}
\left[
\frac{\beta^{2} \tilde{J}^{2}}
{2M^{2}}
\sum_{ij}
\left(
\sum_{k}
\sum_{\alpha}
\sigma_{i}^{\alpha} (k) 
\sigma_{j}^{\alpha} (k)
\right)^{2}
\right] \nonumber \\
\mbox{} & = & 
{\rm tr}_{\{\sigma\}}
{\exp}
\left[
\frac{\beta j_{0}}{M}
\sum_{k}
\sum_{\alpha}
\sum_{ij}
\sigma_{i}^{\alpha}(k)
\sigma_{j}^{\alpha}(k)
+ 
B\sum_{k}\sum_{\alpha}\sum_{i}
\sigma_{i}^{\alpha}(k)
\sigma_{i}^{\alpha}(k+1) 
\right] \nonumber \\
\mbox{} & \times & 
{\exp}
\left[
\frac{\beta^{2} \tilde{J}^{2}}
{2M^{2}}
\sum_{kk^{'}}
\sum_{\alpha \beta}
\sum_{ij}
\sigma_{i}^{\alpha}(k)
\sigma_{j}^{\alpha}(k)
\sigma_{i}^{\beta} (k^{'})
\sigma_{j}^{\beta} (k^{'})
\right] 
\end{eqnarray}
where the bracket was defined as $\ll \cdots \gg = 
\int \prod_{ij}dJ_{ij}
P(J_{ij}) (\cdots)$. 
To take a proper thermodynamic limit, we use the 
scaling 
\begin{eqnarray}
j_{0} & = & 
\frac{J_{0}}{N},\,\,\,\,J   =  \frac{\tilde{J}}{\sqrt{N}}. 
\end{eqnarray}
For this rescaling of the parameters,  
the averaged replicated partition function 
$\ll Z_{M}^{n} \gg$ reads 
\begin{eqnarray}
\ll Z_{M}^{n} \gg  & = & 
{\rm tr}_{\{\sigma\}}
\int _{-\infty}^{\infty}
\prod_{k}
\prod_{\alpha}
\frac{dm_{\alpha}(k)}
{\sqrt{2\pi M/\beta J_{0} N}}
\int_{-\infty}^{\infty}
\prod_{kk^{'}}
\prod_{\alpha \beta}
\frac{dq_{\alpha\beta} (k,k^{'})}
{\sqrt{2\pi M/\beta \tilde{J}\sqrt{N}}} 
\int_{-\infty}^{\infty}
\prod_{kk^{'}}
\prod_{\alpha}
\frac{d\chii_{\alpha\alpha} (k,k^{'})}
{\sqrt{2\pi M/\beta \tilde{J}\sqrt{N}}}  \nonumber \\
\mbox{} & \times & 
{\exp}
\left[
-\frac{\beta J_{0}N}{2M} 
\sum_{k}
\sum_{\alpha}
m_{\alpha} (k)^{2}
-\frac{(\beta \tilde{J})^{2}N}{2M^{2}}
\sum_{kk^{'}}\sum_{\alpha \beta}
q_{\alpha\beta}(k,k^{'})^{2}
-\frac{(\beta \tilde{J})^{2}N}{2M^{2}}
\sum_{kk^{'}}\sum_{\alpha}
\chii_{\alpha\alpha}(k,k^{'})^{2}
\right]  \nonumber \\
\mbox{} & \times & 
{\exp}
{\biggr [}
\frac{\beta J_{0}}
{M} \sum_{k}
\sum_{\alpha}
m_{\alpha}(k)
\sum_{i}
\sigma_{i}^{\alpha} (k)
+
\left(
\frac{\beta \tilde{J}}{M}
\right)^{2}
\sum_{kk^{'}}
\sum_{\alpha \beta}
q_{\alpha\beta} (k,k^{'}) 
\sum_{i}
\sigma_{i}^{\alpha} (k) 
\sigma_{i}^{\beta} (k^{'}) \nonumber \\
\mbox{} & + & 
\left(
\frac{\beta \tilde{J}}{M}
\right)^{2}
\sum_{kk^{'}}
\sum_{\alpha}
\chii_{\alpha \alpha}
(k,k^{'})
\sum_{i}
\sigma_{i}^{\alpha}(k)
\sigma_{i}^{\alpha}(k^{'})
+
B\sum_{k}\sum_{\alpha}\sum_{i}
\sigma_{i}^{\alpha}(k)
\sigma_{i}^{\alpha}(k+1) 
{\biggr ]}  
\end{eqnarray}
We next assume the replica symmetry and 
static approximations such as 
\begin{eqnarray}
m_{\alpha}(k) & = & 
\langle \sigma^{\alpha} (k) \rangle = 
\frac{1}{N}
\sum_{i}\sigma_{i}^{\alpha}(k)=
m \\
q_{\alpha \beta}(k,k^{'}) & = & 
\langle \sigma^{\alpha}(k) \sigma^{\beta} (k^{'}) \rangle = 
\frac{1}{N}\sum_{i}\sigma_{i}^{\alpha}(k) 
\sigma_{i}^{\beta} (k^{'}) = 
q \\
\chii_{\alpha \alpha} (k,k^{'}) & = & 
\langle \sigma^{\alpha} (k) 
\sigma^{\alpha} (k^{'}) \rangle = 
\frac{1}{N}\sum_{i}\sigma_{i}^{\alpha} (k) 
\sigma_{i}^{\alpha}(k^{'}) = 
\chii
\end{eqnarray}
Then, we should notice the relation: 
\begin{eqnarray}
\left(
\frac{\beta \tilde{J}}
{M}
\right)^{2} q 
\sum_{kk^{'}}
\sum_{\alpha \beta} 
\sum_{i}
\sigma_{i}^{\alpha}(k)
\sigma_{i}^{\beta} (k^{'}) & = & 
\left(
\frac{\beta \tilde{J}}
{M}
\right)^{2} q 
\sum_{i}
\left(
\sum_{k}
\sum_{\alpha}
\sigma_{i}^{\alpha} (k)
\right)^{2} - 
\left(
\frac{\beta \tilde{J}}
{M}
\right)^{2} q 
\sum_{i}
\sum_{\alpha}
\left(
\sum_{k}
\sigma_{i}^{\alpha} (k)
\right)^{2} \\
\left(
\frac{\beta \tilde{J}}
{M}
\right)^{2} \chii
\sum_{kk^{'}}
\sum_{\alpha}
\sum_{i}
\sigma_{i}^{\alpha} (k)
\sigma_{i}^{\alpha} (k^{'}) & = & 
\left(
\frac{\beta \tilde{J}}
{M}
\right)^{2} \chii
\sum_{i}
\sum_{\alpha}
\left(
\sum_{k}
\sigma_{i}^{\alpha}(k)
\right)^{2}
\end{eqnarray}
To take into account the above relations, 
we obtain in the limit of $N \to \infty$ as 
\begin{eqnarray}
\ll Z_{M}^{n} \gg  & = & 
\int _{-\infty}^{\infty}
\prod_{k}
\prod_{\alpha}
\frac{dm_{\alpha}(k)}
{\sqrt{2\pi M/\beta J_{0} N}}
\int_{-\infty}^{\infty}
\prod_{kk^{'}}
\prod_{\alpha \beta}
\frac{dq_{\alpha\beta} (k,k^{'})}
{\sqrt{2\pi M/\beta \tilde{J} \sqrt{N}}} 
\int_{-\infty}^{\infty}
\prod_{kk^{'}}
\prod_{\alpha}
\frac{d\chii_{\alpha\alpha} (k,k^{'})}
{\sqrt{2\pi M/\beta \tilde{J} \sqrt{N}}} \nonumber \\
\mbox{} & \times & 
{\exp}
{\biggr [}
nN {\biggr (}
-\frac{\beta J_{0}}{2}  m^{2} + 
\frac{(\beta \tilde{J})^{2}}{4} q^{2}
-\frac{(\beta \tilde{J})^{2}}{4} \chii^{2}  \nonumber \\
\mbox{} & + &  
\int_{-\infty}^{\infty} Dy 
\ln 
\int_{-\infty}^{\infty}
Du\, 
2\cosh \beta \sqrt{
(J_{0}m + \tilde{J}\sqrt{q}y + \tilde{J}\sqrt{\chii - q} u)^{2}
+\Gamma^{2}}
{\biggr )}
{\biggr ]}  \simeq   \exp[nNf].
\end{eqnarray}
Therefore, the following $f$ is regarded as free energy per spin by 
the definition of replica theory
\begin{eqnarray}
f & = & 
-\frac{\beta J_{0}}{2}  m^{2} + 
\frac{(\beta \tilde{J})^{2}}{4} q^{2}
-\frac{(\beta \tilde{J})^{2}}{4} \chii^{2} \nonumber \\
\mbox{} & + &  
\int_{-\infty}^{\infty} Dy 
\ln 
\int_{-\infty}^{\infty}
Du\, 
2\cosh \beta \sqrt{
(J_{0}m + \tilde{J}\sqrt{q}y + \tilde{J} \sqrt{\chii-q} \,u)^{2}
+\Gamma^{2}}
\end{eqnarray}
\subsection{Saddle point equations}
For simplicity, we define 
\begin{eqnarray}
b & = & 
J_{0} m + 
\tilde{J}\sqrt{q}y + 
\tilde{J}\sqrt{\chii-q}u \\
\Theta & = & 
\sqrt{b^{2}+\Gamma^{2}}
\end{eqnarray}
Then, we have the following simplified free energy
\begin{eqnarray}
f & = & 
-\frac{\beta J_{0}}{2}  m^{2} + 
\frac{(\beta \tilde{J})^{2}}{4} (q^{2}-\chii^{2})
+  
\int_{-\infty}^{\infty}
Dy \ln 
\int_{-\infty}^{\infty}
Du \, 2\cosh 
\beta \Theta.
\end{eqnarray}
The saddle point equations are derived as follows
\cite{Thirumalai,Goldschmidt}. 
\begin{eqnarray}
m & = & 
\int_{-\infty}^{\infty}
Dy 
\left[ 
\frac{
\int_{-\infty}^{\infty}
Du \left(
\frac{b}{\Theta}
\right) \sinh \beta \Theta}
{
\int_{-\infty}^{\infty}
Du 
\cosh \beta \Theta}
\right] 
\label{eq:mz3} \\
q & = & 
\int_{-\infty}^{\infty}
Dy 
\left[ 
\frac{
\int_{-\infty}^{\infty}
Du \left(
\frac{b}{\Theta}
\right) \sinh \beta \Theta}
{
\int_{-\infty}^{\infty}
Du 
\cosh \beta \Theta}
\right]^{2}  
\label{eq:q3} \\
\chii & = & 
\int_{-\infty}^{\infty}
Dy 
\left[ 
\frac{
\int_{-\infty}^{\infty}
Du 
\left\{
\left(
\frac{b^{2}}{\Theta^{2}}
\right) \cosh \beta \Theta +
\frac{\Gamma \beta^{-1}}{{\Theta}^{3}} 
\sinh \beta \Theta
\right\}
}
{
\int_{-\infty}^{\infty}
Du 
\cosh \beta \Theta}
\right]  
\label{eq:chi3} \\
m_{x} & = & 
\frac{\partial f}{\partial \Gamma} = 
\int_{-\infty}^{\infty}
Dy 
\left[ 
\frac{
\int_{-\infty}^{\infty}
Du \left(
\frac{\Gamma}{\Theta}
\right) \sinh \beta \Theta}
{
\int_{-\infty}^{\infty}
Du 
\cosh \beta \Theta}
\right]
\label{eq:mx3}
\end{eqnarray}
\subsection{At the ground state}
We first should notice that 
$\tilde{q}$ is always larger than $q$. 
In fact, we can easily show that 
\begin{eqnarray}
\chii & = & 
\int_{-\infty}^{\infty}
Dy 
\left[ 
\frac{
\int_{-\infty}^{\infty}
Du 
\left(
\frac{b^{2}}{\Theta^{2}}
\right) \cosh \beta \Theta
}
{
\int_{-\infty}^{\infty}
Du 
\cosh \beta \Theta}
\right]   \nonumber \\
\mbox{} & \geq & 
\int_{-\infty}^{\infty}
Dy 
\left[ 
\frac{
\int_{-\infty}^{\infty}
Du 
\left(
\frac{b^{2}}{\Theta^{2}}
\right) \sinh \beta \Theta
}
{
\int_{-\infty}^{\infty}
Du 
\cosh \beta \Theta}
\right]   \nonumber \\
\mbox{} & \geq & 
\int_{-\infty}^{\infty}
Dy 
\left[ 
\frac{
\int_{-\infty}^{\infty}
Du 
\left(
\frac{b^{2}}{\Theta^{2}}
\right) \sinh \beta \Theta
}
{
\int_{-\infty}^{\infty}
Du 
\cosh \beta \Theta}
\right]^{2} =  q.  
\end{eqnarray}
Then, we consider the limit of $\beta \to  \infty$. 
If $\chii -q=\epsilon \geq 0$ 
is of order $1$ object, the free energy $f$ diverges 
in the limit of $\beta \to \infty$ 
as $(\beta \tilde{J})^{2} (q^{2}-\chii^{2})/4$. 
Therefore, we conclude that 
$q = \chii$ should be satisfied in the limit of 
$\beta \to \infty$ and we obtain the saddle pint equation at the ground state as 
\begin{eqnarray}
m & = & 
\int_{-\infty}^{\infty}
Dy 
\left(
\frac{b}{\Theta}
\right)  = 
\int_{-\infty}^{\infty}
Dy 
\frac{(J_{0}m+\tilde{J}\sqrt{q}y)}
{\sqrt{(J_{0}m+\tilde{J}\sqrt{q}y)^{2}+\Gamma^{2}}} \\
q & = &  \chii = 
\int_{-\infty}^{\infty}
Dy 
\left(
\frac{b}{\Theta}
\right)^{2}  = 
\int_{-\infty}^{\infty}
Dy 
\left\{
\frac{(J_{0}m+\tilde{J}\sqrt{q}y)}
{\sqrt{(J_{0}m+\tilde{J}\sqrt{q}y)^{2}+\Gamma^{2}}} 
\right\}^{2} \\
m_{x} & = & 
\int_{-\infty}^{\infty}
Dy 
\left(
\frac{\Gamma}{\Theta}
\right)  = 
\int_{-\infty}^{\infty}
Dy 
\frac{\Gamma}
{\sqrt{(J_{0}m+\tilde{J}\sqrt{q}y)^{2}+\Gamma^{2}}}. 
\end{eqnarray}

\section{Exact analysis at T = 0}

\noindent First, let us consider the case of pure LRIAF model and
  rewrite our Hamiltonian
{$H$} in Eq.(1) for
$\tilde{J}=0$ as
\begin{equation}
H =  \frac{1}{2N} \left(
{ \sum_{i=1}^N \sigma^z_i}
\right)^2  -  \frac{1}{N} \sum_{i=1}^N {(\sigma^z_i)}^2 - h \sum_{i=1}^N \sigma^z_i - \Gamma \sum_{i=1}^N \sigma^x_i 
\end{equation}
If we now denote the total spin by $\vec{\sigma}_{tot}$ i.e.
$\vec{\sigma}_{tot} = \frac{1}{N} \sum_{i=1}^N \vec{\sigma}_i$
(where $N|\vec{\sigma}|= 0,1,2,....,N$), then the Hamiltonian
{$H$} can be expressed as
\begin{equation}
\frac{H}{N} =  \frac{1}{2} {(\sigma^z_{tot})}^2 - h \sigma^z_{tot} - \Gamma \sigma^x_{tot} - \frac{1}{N} .
\end{equation}
Let us assume the average total spin $\langle\vec{\sigma}\rangle$ to be oriented at an angle $\theta$ with
the $z$-direction : $\langle\sigma^z_{tot}\rangle = m \cos \theta$ and
$\langle\sigma^x_{tot}\rangle = m \sin \theta$. Hence the average
total energy $E_{tot} = \langle H \rangle$ can be written as
\begin{equation}
\frac{E_{tot}}{N} =  \frac{1}{2} m^2 {{\cos}^2 \theta} - h m {\cos \theta} - \Gamma m  \sin \theta - \frac{1}{N} .
\end{equation}
At the zero temperature and at $\Gamma = 0$, for $h = 0$, the energy
$E_{tot}$ is
minimised when {$\theta = 0$} and {$m = 0$} (complete antiferromagnetic
order in $z$-direction). As soon as $\Gamma \ne 0$ ($h = 0$) the minimisation
of $E_{tot}$ requires {$\theta = \pi/2$} and $m = 1$ (the maximum possible
value); driving the system to paramagnetic phase. This discontinuous transition
at $T = 0$ was also seen in \cite{Vidal}.
As observed in our Monte Carlo study in the previous
section, $\Gamma_c(T) \rightarrow 0$ as
$T \rightarrow 0$. This is consistent with this exact result $\Gamma_c = 0$
at $T = 0$. For $T = 0$ (and $h = 0$), therefore, the transition from
antiferromagnetic ($\theta = 0 = m$) to para ($\theta = \pi/2, m = 1$)
phase, driven by the transverse field $\Gamma$, occurs at $\Gamma = 0$ itself.

One can also
calculate the susceptibility $\chi$ at $\Gamma = 0 = T$. Here
$E_{tot}/N = \frac{1}{2} m^2 {{\cos}^2 \theta} - h m {\cos \theta} - \frac{1}{N}$ and the minimisation of
this energy gives $m \cos \theta = h$ giving the (longitudinal)
susceptibility $\chi = {m \cos \theta}/{h} = 1$. This is consistent with
the observed behaviour of $\chi$ shown in Fig. \ref{fig:inset5} where the
extrapolated value of
$\chi$ at $\Gamma = 0$ increases with decreasing $T$ and approaches $\chi = 1$
as $T \rightarrow 0$.

At finite temperatures $T \ne 0$, for $h = 0$, we have to consider also the
entropy term
and minimise the free energy ${\mathcal F} = E_{tot} - TS$ rather than
$E_{tot}$ where $S$ denotes the entropy of the state.
This entropy term will also take part in fixing the value of $\theta$ and
$m$ at which the free energy ${\mathcal F}$ is minimised. As soon
as the temperature $T$ becomes non-zero, the extensive entropy of the system
for
antiferromagnetically ordered state with $m \simeq 0$ (around and close-by
excited states with $\theta = 0$) helps stabilisation
near $\theta = 0$ and $m = 0$ rather than near the para phase with
$\theta = \pi/2$ and $m = 1$, where the entropy drops to zero. While the
transverse field tends to align the spins along $x$
direction (inducing $\theta = \pi/2$ and $m = 1$), the entropy factor
prohibits that and the system adjusts $\theta$ and
$m$ values accordingly and they do not take the disordered or para state
values
($\theta = \pi/2$ and $m = 1$) for any non-zero value of $\Gamma$ (like
at $T = 0$).
For very large values of $\Gamma$, of
course, the free energy ${\mathcal F}$ is practically dominated by the
transverse field term in $H$ and again {$\theta = \pi/2$}
and $m = 1$, beyond $\Gamma = \Gamma_c(T) > 0$ for $T > 0$.
However, this continuous transition-like behaviour may be argued
\cite{Diptiman} to correspond to a
crossover type property of the model at finite temperatures (suggesting
that the observed finite values of $\Gamma_c(T)$ are only effective numerical
values). In fact, for $h = 0$ one adds the entropy term $-T\ln D$,
where $D$ is the degeneracy for getting total spin 
$\tau = N|\vec{\sigma}_{tot}|$ \cite{Vidal},
\be
D = \frac{N!}{(N/2 + \tau)!(N/2 - \tau)!} -
\frac{N!}{(N/2 + \tau + 1)!(N/2 - \tau - 1)!}
\ee 
to $E_{tot}$ in Eq.(A3) to get ${\mathcal F}$ and
one can then get, after minimising the ${\mathcal F}$ with
respect to $m$ and $\theta$, $m = \tanh (\Gamma/2T)$, which
indicates an analytic variation of $m$
and no phase transition
at any finite temperature for $\tilde{J}=0$
(antiferromagnetic phase
occurs only at
$\Gamma=T=0$ and $J_0/\tilde{J} = \infty$ as shown in
Fig. \ref{fig:fg2}).


\end{document}